\documentclass{phc-proc4-auth}
\usepackage{graphicx,dcolumn,bm,amssymb}
\begin{document}

\begin{frontmatter}
\title{ Multiband superconductivity in NbSe$_2$ from heat transport}

\author[TOR]{E.~Boaknin},
\author[TOR]{M.~A.~Tanatar\thanksref{MAT}},
\author[TOR]{J.~Paglione},
\author[TOR]{D.~G.~Hawthorn},
\author[TOR]{R.~W.~Hill},
\author[TOR]{F.~Ronning},
\author[TOR]{M.~Sutherland},
\author[TOR]{L.~Taillefer\corauthref{COR}\thanksref{LT}},
\author[SF]{ J.~Sonier},
\author[Bristol]{ S.~M.~Hayden},
\author[UKy]{J.~W.~Brill}

\corauth[COR]{Corresponding author:
Louis.Taillefer@physique.usherb.ca}
\address[TOR]{Department of Physics, University of Toronto, Toronto, Canada M5S 1A7}
\address[SF]{Department of Physics, Simon Fraser University, Burnaby, Canada V5A 1S6}
\address[Bristol] {H. H. Wills Physics Laboratory, University of Bristol, United Kingdom}
\address[UKy]{Department of Physics and Astronomy, University of Kentucky, Lexington, Kentucky, USA 40506-0055}

\thanks[MAT]{Permanent address: Inst. Surf. Chem., N.A.S. Ukraine.}
\thanks[LT]{Current address: Department of Physics, University of Sherbrooke, Sherbrooke, Canada J1K 2R1}

\begin{abstract}
The thermal conductivity of the layered $s$-wave superconductor
NbSe$_2$ was measured down to $T_c$/100 throughout the vortex
state.  With increasing field, we identify two regimes: one with
localized states at fields very near $H_{c1}$ and one with highly
delocalized quasiparticle excitations at higher fields. The two
associated length scales are most naturally explained as
multi-band superconductivity, with distinct small and large
superconducting gaps on different sheets of the Fermi surface.
\end{abstract}

\end{frontmatter}


In the classical theory of superconductivity all electrons on the
Fermi surface contribute equally to the superconducting pairing,
giving a constant superconducting gap $\Delta$. The difference of
$\Delta$ on different sheets of the Fermi surface, or multiband
superconductivity (MBSC), considered theoretically back in the 50s
\cite{Mathias}, has emerged recently as a possible explanation for
the unusual properties of MgB$_2$ \cite{MgB2}.

Based on the observation of a sizable difference of $\Delta$ on
two Fermi surface sheets by angle resolved photoemission
spectroscopy, it has been suggested that the layered
superconductor NbSe$_2$ is also host to MBSC \cite{yokoya}.
However, these surface-sensitive measurements were performed only
down to 5.3 K, close to $T_c$=7.0 K. We present evidence of bulk
MBSC at low temperatures in this compound.


The thermal conductivity $\kappa$ of pure samples of NbSe$_2$
(residual resistivity $\rho_0$=3 $\mu \Omega$~cm) was measured
upon warming in a magnetic field ($H \parallel c$ axis)
\cite{Boaknin}.
In Fig.~1, $\kappa$ is presented as $\kappa/T$ vs $T^2$, enabling
a separation of the electronic contribution, $\kappa_0$ $\sim T$,
and the phononic contribution, $\kappa ^g \sim T^3$. In zero
field, $\kappa_0/T$ goes to zero in the $T$=0 limit as expected
for $s$-wave superconductors, thus ruling out the possibility of
nodes in the gap at any point on the Fermi surface. With the
application of a magnetic field, $\kappa_0/T$ rapidly increases
(Fig.~\ref{fig:2}, main panel) up to $H_{c2}$, following closely
the increase of the electronic specific heat $\gamma$
\cite{SpecheatNbSe}. This behavior is very different from that
expected for conventional superconductors. There, electronic
quasiparticles remain localized inside the vortex cores, such that
the increase in quasiparticle density (seen as an increase of
$\gamma$ with $H$) is not followed by an increase of $\kappa_0$.
This is shown for V$_3$Si in the upper panel of Fig.~2
\cite{SpecheatV3Si,ramirez}.

\begin{figure}
 \centering
  \includegraphics[totalheight=2.55in]{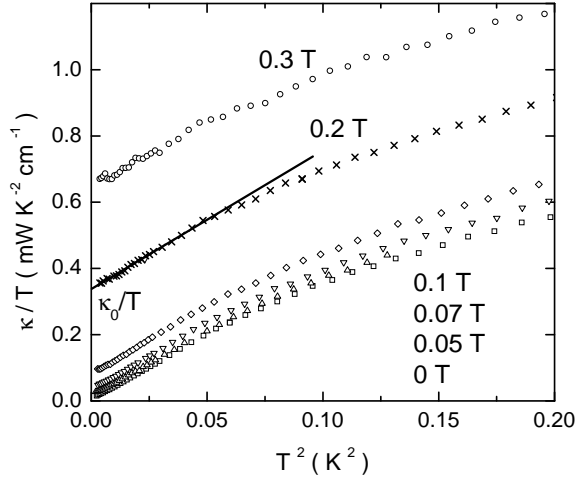}
 \caption{\label{fig:1} $\kappa/T$ vs $T^2$ for NbSe$_2$. The line shows the $T \to 0$ extrapolation which yields the
 electronic contribution $\kappa_0/T$.
 In zero field, $\kappa_0/T$=0, as expected for a superconductor without nodes in the gap anywhere
 on the Fermi surface.}
\end{figure}

A closer examination of the field dependence of $\kappa_0$ and of
$\gamma$ at fields close to $H_{c1}$ is consistent with the
presence of localized states, another indication of a gap without
nodes.
Indeed, in a limited field range, $\kappa_0$ shows an activated
increase in field whereas the specific heat increases rapidly
above $H_{c1}$ (lower inset in Fig.~\ref{fig:2}).

\begin{figure}
 \centering
 \includegraphics[totalheight=2.55in]{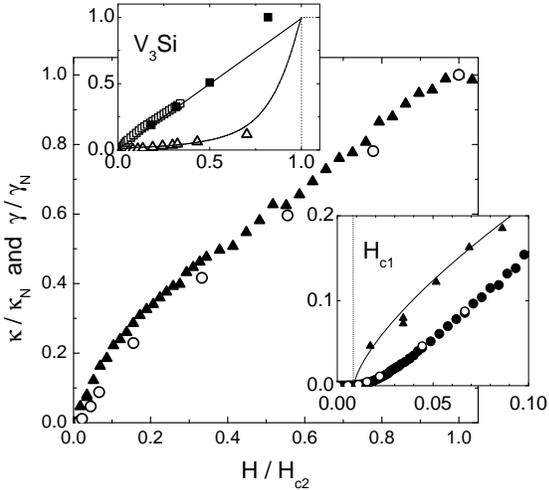}
\caption{\label{fig:2} Field dependence of thermal conductivity
(circles for NbSe$_2$ and empty triangles for V$_3$Si) and
specific heat (filled triangles for NbSe$_2$ and squares for
V$_3$Si) in NbSe$_2$ (main panel and lower inset) and V$_3$Si
(upper inset). The filled circles come from a field sweep at
100~mK. Both quantities are normalized to the values in the normal
state. }
\end{figure}

\begin{figure}
 \centering
 \includegraphics[totalheight=2.55in]{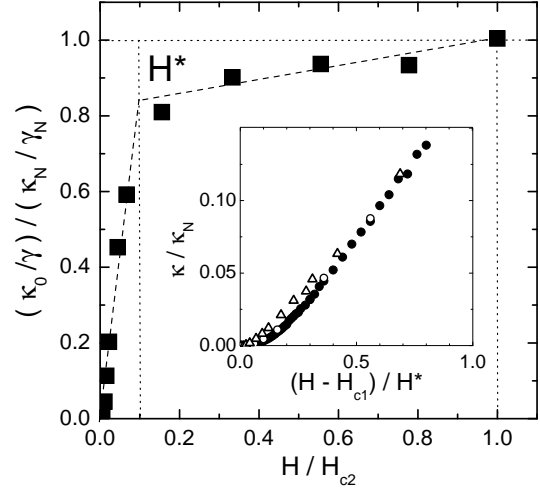}
 \caption{\label{fig:3} Field dependence of $\kappa_0 / \gamma$ (main panel).
 The inset shows the field dependence of thermal conductivity for NbSe$_2$ (circles) and V$_3$Si (triangles),
 plotted as a function of $(H-H_{c1})/H^*$ where $H^*=H_{c2}/9$
 for NbSe$_2$ and $H^*=H_{c2}$ for V$_3$Si.
}
\end{figure}

In Fig.~\ref{fig:3}, we show a more detailed comparison of
$\kappa$ and $\gamma$ by plotting the ratio $\kappa _0 / \gamma$.
In a rough sense, it represents the "mobility" of quasiparticle
excitations in the vortex state. It is clear that the field
dependence of $\kappa _0 / \gamma$ shows a strong change of slope
at $H^* \sim H_{c2}/9$, supporting the existence of a field scale
which plays the same role for quasiparticle localization as
$H_{c2}$ does in standard $s$-wave superconductors. In the inset
of Fig.~\ref{fig:3}, we show that the low field behavior of
NbSe$_2$ can be matched to that of V$_3$Si by using $H^* \sim
H_{c2}$/9.

While $\xi$ is associated with the upper critical field ($\xi^2 =
\Phi _0 /2 \pi H_{c2}$), a second length scale $\xi ^*$ must be
associated with $H^*$.
%
The crossover between $\xi$ and $\xi ^*$
with $H$ can naturally explain the shrinking of the vortex cores
observed by muon spin relaxation in NbSe$_2$ \cite{muSR}.
Moreover, the superconducting coherence length is is related to
$\Delta$ via $\xi \sim v_F/\Delta$, where $v_F$ is the Fermi
velocity. In NbSe$_2$, $v_F$ is approximately the same for
different sheets of the Fermi surface, such that the ratio $\xi^*
/\xi ~ \sim$~3 gives a ratio of associated superconducting gaps
$\Delta /\Delta^* \sim$~3. This value is consistent with previous
heat capacity and tunneling results
\cite{SpecheatNbSe,sanchez,hess}.


In conclusion, we have identified the anomalous evolution of
thermal conductivity in NbSe$_2$ versus magnetic field with the
existence of two length scales in the superconducting state. This
finds a natural explanation in a model of multi-band
superconductivity, assuming a ratio of larger gap to smaller gap
of about 3.


This work was supported by the Canadian Institute for Advanced Research and funded by NSERC.


\end{document}